\documentclass[options]{JHEP3}
\usepackage{epsfig}

\title{Density matrix renormalization group\\
in a two-dimensional $\lambda\phi^4$ Hamiltonian lattice model}
\author{Takanori Sugihara\\
RIKEN BNL Research Center, \\
Brookhaven National Laboratory, 
Upton, New York 11973, USA\\ E-mail: \email{sugihara@bnl.gov}}
\abstract{
Density matrix renormalization group (DMRG) is applied to 
a (1+1)-dimensional $\lambda\phi^4$ model. 
Spontaneous breakdown of discrete $Z_2$ symmetry is studied numerically 
using vacuum wavefunctions. We obtain the critical coupling 
$(\lambda/\mu^2)_{\rm c}=59.89\pm 0.01$ 
and the critical exponent $\beta=0.1264\pm 0.0073$, which are 
consistent with the Monte Carlo and the exact results, respectively. 
The results are based on extrapolation to the continuum limit 
with lattice sizes $L=250,500$, and $1000$. 
We show that the lattice size $L=500$ is sufficiently 
close to the the limit $L\to\infty$. 
}


\preprint{}

\begin{document}

\section{Introduction}
Hamiltonian diagonalization is a useful method 
for nonperturbative analysis of many-body 
quantum systems \cite{Bethe:1931hc,Yang:1967bm}. 
If Hamiltonian is diagonalized, the system can be analyzed 
nonperturbatively at the amplitude level using the obtained 
wave functions 
\cite{Harada:1993va,Sugihara:xq,Sugihara:1997xh,
Sugihara:2001ch,Sugihara:2001ci}. 
In addition, one can discuss associated symmetry 
based on operator algebra \cite{Sakai:1983dg,Fendley:2002sg}. 
However, in general quantum field theories, 
the method does not work without reducing degrees of freedom 
because the dimension of Hamiltonian increases exponentially
as the system size becomes large. 
There is a severe limitation on numerical diagonalization of Hamiltonian. 
To apply the method to quantum field theories, 
we need to find a way to remove unimportant degrees of 
freedom and create a small number of optimum basis states. 
This is the concept of renormalization group. 

Wilson's numerical renormalization group applied to 
the Kondo problem was a successful consideration 
along this line \cite{Wilson:1974mb}. 
To analyze chain models other than Kondo Hamiltonian, 
White proposed density matrix renormalization group (DMRG) 
as an extension of the Wilson's method \cite{white1,white2}. 
In DMRG, calculation accuracy of the target state can be 
controlled systematically using density matrices. 
White calculated energy spectra and wavefunctions of Heisenberg chains 
composed of more than 100 sites using a standard workstation. 
The DMRG analysis of a 100-site $S=1/2$ chain corresponds to 
diagonalization of Hamiltonian with 
$2^{100}\sim 10^{30}$ dimensions. 
DMRG has been applied to various one-dimensional models, 
such as Kondo, Hubbard, and $t$-$J$ chain models, 
and achieved great success. 
A two-dimensional Hubbard model has also been studied with DMRG 
in both real- and momentum-space representation \cite{xiang,xls}. 
DMRG works well on small two-dimensional lattices 
and new techniques have been proposed for larger lattices. 
DMRG has also been extended to finite-temperature 
chain models using the transfer-matrix technique 
based on the Suzuki-Trotter formula 
\cite{ft,wx,xiang2,shiba,dmrglec}. 
In particle physics,  the massive Schwinger model has been 
studied using DMRG to confirm the well-known Coleman's picture 
of `half-asymptotic' particles 
at a background field $\theta=\pi$ \cite{Byrnes:2002nv}.
It is interesting to seek a possibility of applying 
the method to quantum chromodynamics (QCD) in order 
to study color confinement and spontaneous chiral symmetry breaking 
based on QCD vacuum wavefunctions. 

In fermionic lattice models, the number of particles 
contained in each site is limited because of the Pauli principle. 
On the other hand, in bosonic lattice models, each site can contain 
infinite number of particles in principle. 
It is not evident whether Hilbert space can be described 
appropriately with a finite set of basis states in bosonic models.
This point becomes crucial when DMRG is applied to 
gauge theories because gauge particles are bosons. 
Before working in lattice gauge theories 
like Kogut-Susskind Hamiltonian \cite{ks}, 
we need to test DMRG in a simple bosonic model 
and recognize how many basis states are necessary for each site 
to reproduce accurate results. 
In this paper, we apply DMRG to a $\lambda\phi^4$ model 
with (1+1) space-time dimensions. 
We define a Hamiltonian model on a spatial lattice 
(only space is discretized) 
because DMRG is a method based on Hamiltonian formalism. 
The model has spontaneous breakdown of discrete $Z_2$ symmetry 
and the exact values of the critical exponents are known. 
We are going to justify the relevance of DMRG truncation 
of Hilbert space in the bosonic model 
by comparing our numerical results with the Monte Carlo 
and the exact results \cite{Loinaz:1997az,onsager}. 
Our largest lattice size is $L=1000$, which is about twice 
of the latest Monte Carlo one \cite{Loinaz:1997az}. 
It is shown that the lattice size $L=500$ is 
sufficiently close to the limit $L\to \infty$. 

This paper is organized as follows. 
In Sec. \ref{lattice}, Hamiltonian lattice formulation 
of the model is given. 
The canonical variables are transformed to Fock-like operators, 
each of which creates or annihilates a boson on each site. 
Real-space representation is maintained 
because local interactions are useful for DMRG. 
In Sec. \ref{dmrg}, DMRG setup for the model is explained. 
A superblock is composed of two blocks and one site. 
Sec. \ref{result} gives the definition of the critical values 
and shows numerical results. 
Sec. \ref{summary} is devoted to summary.

\section{A $\lambda\phi^4_{1+1}$ Hamiltonian lattice model}
\label{lattice}
We start from the following Lagrangian 
of a (1+1)-dimensional $\lambda\phi^4$ model 
\begin{equation}
  L_{\rm c} = \int dx
  \left[
  \frac{1}{2}(\partial_\mu\phi \partial^\mu\phi-\mu_0^2\phi^2)
  -\frac{\lambda}{4!}\phi^4
  \right], 
\end{equation}
where space and time are continuous. 
Only the space is discretized to obtain a Hamiltonian lattice model. 
In this paper, the spatial derivative is modeled 
as a naive difference on the lattice.
\footnote{Discretization errors 
associated with the derivative can be improved in a systematic way 
by introducing non-nearest neighbor interactions
\cite{Sugihara:2003mh,Sugihara:2003ga}.}
Lagrangian of the lattice model is
\begin{equation}
  \displaystyle L = a\sum_{n=1}^L
  \left(
    \frac{1}{2}\dot{\phi}_n^2 - \frac{\mu_0^2}{2}\phi_n^2 
    -\frac{\lambda}{4!}\phi_n^4
  \right)
  \displaystyle -\frac{1}{2a}\sum_{n=1}^{L-1}
    \left(\phi_n-\phi_{n+1}\right)^2, 
\end{equation}
where $a$ is a lattice spacing and time is continuous. 
Open boundary conditions are chosen. 
Hamiltonian is given by 
\begin{equation}
  \displaystyle H = \sum_{n=1}^L
  \left(
    \frac{1}{2a}\pi_n^2 + \frac{\mu_0^2 a}{2}\phi_n^2
    +\frac{\lambda a}{4!}\phi_n^4
  \right)
  \displaystyle +\frac{1}{2a}\sum_{n=1}^{L-1}
    \left(\phi_n-\phi_{n+1}\right)^2, 
  \label{ham}
\end{equation}
where $\pi_n\equiv a\dot{\phi}_n$ is the conjugate variable 
to $\phi_n$. 
All dimensionful quantities are measured in units of $a$. 
\begin{equation}
  \tilde{H} \equiv Ha, 
  \quad
  \tilde{\mu_0}^2 \equiv \mu_0^2 a^2, 
  \quad
  \tilde{\lambda} \equiv \lambda a^2. 
\end{equation}
To quantize the system, 
we impose an equal-time commutation relation 
\begin{equation}
    [\phi_m(t),\pi_n(t)] = i\delta_{mn}. 
\end{equation}
We rewrite Hamiltonian (\ref{ham}) using
creation and annihilation operators $a_n^\dagger$ and $a_n$. 
\begin{equation}
  \phi_n = \frac{1}{\sqrt{2}}
    \left(a_n^\dagger + a_n\right), \quad
  \pi_n = \frac{i}{\sqrt{2}}
    \left(a_n^\dagger - a_n\right), 
  \label{ca}
\end{equation}
where $[a_m,a_n^\dagger]=\delta_{mn}$ and $a_n|0\rangle=0$. 
Note that $a_n^\dagger$ and $a_n$ are not creation 
and annihilation operators in Fock representation. 
The index $n$ of the operators $a_n^\dagger$ and $a_n$ 
stands for the discretized spatial coordinate, not momentum. 
Real-space representation is better for our purpose 
because local interactions are useful for DMRG. 
(See Ref. \cite{xiang,xls} 
for DMRG in momentum-space representation.)

\FIGURE{
 \epsfig{file=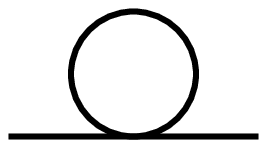,width=4.0cm}
\caption{The only divergent diagram in the (1+1)-dimensional 
$\lambda\phi^4$ model. }
\label{selfmass}}

In this model, Fig. \ref{selfmass} is the only diagram 
that diverges in the continuum limit $a\to 0$. 
The divergence can be renormalized by redefining 
the mass parameter as 
\[
  \tilde{\mu}_0^2 = \tilde{\mu}^2 - \delta\tilde{\mu}^2, 
\]
where $\delta\tilde{\mu}^2$ is a counter term that 
cancels the divergence. 
\[
  \delta\tilde{\mu}^2 = \frac{\lambda}{2}S_L(\tilde{\mu}^2). 
\]
$S_L$ is given as a discrete sum 
because the system is quantized in a finite spatial box, 
\begin{equation}
  \displaystyle
  S_L(\tilde{\mu}^2) = \frac{1}{2L}\sum_{n=1}^{L}
  \frac{1}{\sqrt{\tilde{\mu}^2 + 4\sin^2 \frac{\pi n}{L}}}. 
  \label{sl}
\end{equation}
In the limit $L\to\infty$, the sum (\ref{sl}) becomes 
\begin{eqnarray}
  S_\infty(\tilde{\mu}^2) &=& \frac{1}{2\pi}\int_0^1 dt
  \frac{1}{\sqrt{(t^2+\frac{\tilde{\mu}^2}{4})(1-t^2)}}
  \nonumber
  \\
  &=&
  \frac{1}{\pi\sqrt{\tilde{\mu}^2+4}}
  F\left(\frac{\pi}{2},\frac{2}{\sqrt{\tilde{\mu}^2+4}}\right),
  \label{si}
\end{eqnarray}
where $F$ is the elliptic integral of the first kind
\[
  F(\varphi,k)
  = \int_0^\varphi \frac{d\theta}{\sqrt{1-k^2\sin^2 \theta}}.
\]
The integral $S_\infty$ is used for all calculations of the 
counter term $\delta\tilde{\mu}^2$ even with a finite lattice 
because we are interested in the limit $L\to\infty$. 
We are going to calculate the critical coupling and exponent 
by extrapolating numerical data points to the limit $L\to\infty$. 

In Fock representation, the divergence can be removed easily 
by taking normal ordering of Hamiltonian. 
However, 
Hamiltonian  (\ref{ham}) cannot be normal-ordered easily 
because our representation is given in real space 
and different from Fock one. 
For this reason, we need to evaluate the integral (\ref{si}) 
explicitly and redefine the mass parameter to renormalize 
the divergence in a numerical manner.

\section{DMRG with one-site insertion}
\label{dmrg}
We are going to apply DMRG technique to the the model given 
in the previous section. 
In this paper, the system is composed of two renormalized blocks 
(system and environment blocks) and one bare site, 
each of which is approximated with a finite number of basis states. 
Based on the DMRG technique, basis states for system and environment 
blocks are optimized to describe the whole system 
in a finite dimensional space. 

In fermionic models, dimension of Hamiltonian is finite 
on a finite lattice. On the other hand, 
in bosonic models, each site can contain any number of bosons. 
Namely, each site has infinite degrees of freedom. 
The dimension of a bosonic Hamiltonian is infinite 
even on a finite lattice. 
There is an essential difference between boson and fermion. 
To perform numerical calculations, Hamiltonian needs to be 
finite dimensional because a computer can only take care of 
finite dimensional matrices. 
For this reason, in our bosonic model, 
the number of basis states of each bare site needs to be restricted 
when it is inserted between system and environment blocks. 
In the White's first paper for DMRG, 
two sites are inserted between system and environment blocks 
in each RG step \cite{white1,white2}. 
Since the model considered there is antiferromagnetic 
$S=1/2$ Heisenberg chain, the total spin of the whole system 
needs to be kept constant for numerical stability. 
However, our model has no such a requirement associated with spins. 
When we insert bare sites between two blocks, 
a smaller number of sites is better. 
One-site insertion is the best choice. 

\FIGURE{
 \epsfig{file=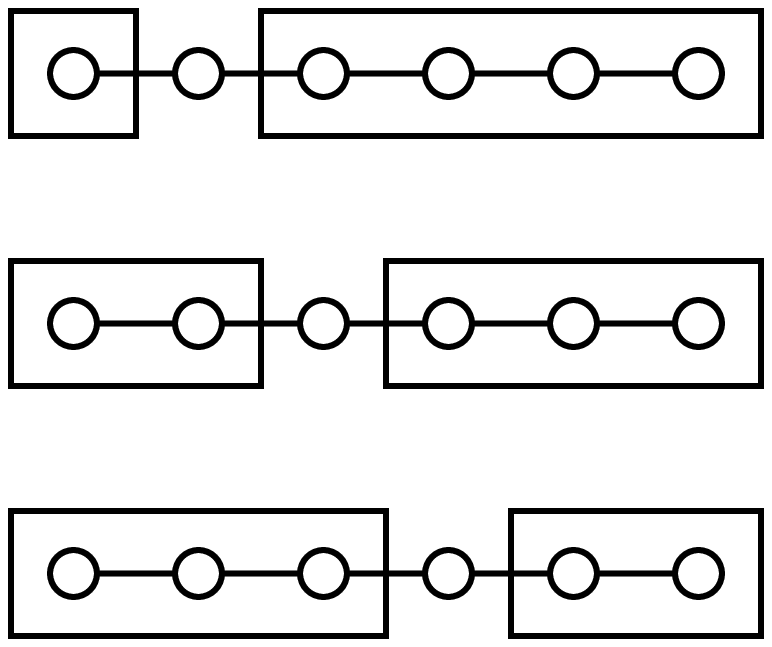,width=5.0cm}
\caption{DMRG sweep process is shown. 
A superblock is composed of two renormalized blocks 
(system and environment) and one bare site. 
A bare site is moved from end to end. 
Basis states are optimized each time. 
One DMRG sweep is composed of $L$ basis optimization steps. }
\label{sweep}}

We divide Hamiltonian into two parts: 
\begin{equation}
  \tilde{H} = \sum_{n=1}^L h_n + \sum_{n=1}^{L-1} h_{n,n+1}, 
\end{equation}
where
\begin{eqnarray*}
  h_n &=& \frac{1}{2}\pi^2_n + \frac{\tilde{\mu}_0^2}{2}\phi_n^2
  +\frac{\tilde{\lambda}}{4!}\phi_n^4, 
  \nonumber
  \\
  h_{n,n+1} &=& \frac{1}{2}(\phi_n-\phi_{n+1})^2. 
\end{eqnarray*}
We are going to apply the finite system algorithm of DMRG 
to the Hamiltonian. 
As shown in Fig. \ref{sweep}, a superblock Hamiltonian $H_{\rm S}$ 
is composed of two blocks and one site: 
\begin{equation}
  H_{\rm S} = \bar{H}_{\rm L} + h_{n-1,n} + h_n + h_{n,n+1}
  + \bar{H}_{\rm R}, 
\end{equation}
where $\bar{H}_{\rm L}$ and $\bar{H}_{\rm R}$ are 
effective Hamiltonian for the left and right blocks, respectively. 
$h_{n-1,n}$ ($h_{n,n+1}$) is an interaction 
between the left (right) block and the inserted $n$-th bare site. 
The superblock Hamiltonian $H_{\rm S}$ is an effective 
Hamiltonian for the original $\tilde{H}$. 

Basis states for the superblock are 
\begin{equation}
  |i,j,k\rangle_n = |u_i^{({\rm L})}\rangle 
  |j^{(n)}\rangle |v_k^{({\rm R})}\rangle, 
\end{equation}
where 
$\{|u_i^{({\rm L})}\rangle|i=1,\dots,M\}$ and 
$\{|v_k^{({\rm R})}\rangle|k=1,\dots,M\}$ are finite sets 
of basis states for the left and right blocks, respectively. 
$\{|j^{(n)}\rangle|j=1,\dots,N\}$ is a set of basis states 
for the $n$-th bare site inserted between the two blocks, 
\begin{equation}
  |j^{(n)}\rangle
  \equiv \frac{1}{\sqrt{(j-1)!}}(a_n^\dagger)^{j-1}|0\rangle, 
  \quad
  j=1,...,N. 
  \label{state}
\end{equation}
At every DMRG step, basis states 
$\{|u_i^{({\rm L})}\rangle\}$ and $\{|v_k^{({\rm R})}\rangle\}$ 
are updated using the density matrix technique to 
renormalize the left and right block Hamiltonian, 
$\bar{H}_{\rm L}$ and $\bar{H}_{\rm R}$. 
(See Appendix \ref{appa} for the creation of basis states.) 
If $M$ and $N$ are infinite, the superblock Hamiltonian 
$H_{\rm S}$ gives exact spectra because update of basis states 
is nothing but infinite dimensional unitary transformation. 
However, such a calculation is not possible on a finite computer. 
We truncate Hilbert space with finite $M$ and $N$. 
The target state is expanded as 
\begin{equation}
  |\Psi\rangle = \sum_{i=1}^M \sum_{j=1}^N \sum_{k=1}^M
  \Psi_{ijk}|i,j,k\rangle_n. 
\end{equation}
The dimension of the superblock Hamiltonian is $M^2N$. 
The relevance of truncation will be checked numerically 
by seeing convergence of energy and wavefunction 
with respect to the parameters $M$ and $N$. 

Matrix elements of the superblock Hamiltonian are 
\begin{eqnarray}
  _n\langle i,j,k | H_{\rm S} |i',j',k'\rangle_n
  =&&
  \langle u_i^{({\rm L})}|\bar{H}_{\rm L}|u_{i'}^{({\rm L})}\rangle
  \delta_{jj'}\delta_{kk'}
  +\langle j^{(n)}|\langle u_i^{({\rm L})} |h_{n-1,n}|
  u_{i'}^{({\rm L})} \rangle|j'^{(n)}\rangle \delta_{kk'}
  \nonumber
  \\
  &&+\delta_{ii'}\langle j^{(n)}|h_n|j'^{(n)}\rangle \delta_{kk'}
  \nonumber
  \\
  &&+ \delta_{ii'}\langle j^{(n)}|\langle v_k^{({\rm R})} |h_{n,n+1}|
  v_{k'}^{({\rm R})} \rangle|j'^{(n)}\rangle
  + \delta_{ii'}\delta_{jj'}
  \langle v_k^{({\rm R})}|\bar{H}_{\rm R}|v_{k'}^{({\rm R})}\rangle, 
\label{h}
\end{eqnarray}
where
\begin{eqnarray*}
  &&\langle j^{(n)}|\langle u_i^{({\rm L})} |h_{n-1,n}|
  u_{i'}^{({\rm L})} \rangle|j'^{(n)}\rangle
  \\
  =&&
  \frac{1}{2}\langle u_i^{({\rm L})}
    |\phi_{n-1}^2|u_{i'}^{({\rm L})}\rangle\delta_{jj'}
  -\langle u_i^{({\rm L})}|\phi_{n-1}
    |u_{i'}^{({\rm L})}\rangle
    \langle j^{(n)}|\phi_n|j'^{(n)}\rangle
  +\frac{1}{2}\langle j^{(n)}|\phi_n^2|j'^{(n)}\rangle\delta_{ii'}, 
\end{eqnarray*}
and
\begin{eqnarray*}
  &&\langle j^{(n)}|\langle v_k^{({\rm R})} |h_{n,n+1}|
  v_{k'}^{({\rm R})} \rangle|j'^{(n)}\rangle 
  \\
  =&&
  \frac{1}{2}\langle v_k^{({\rm R})} |\phi_{n+1}^2
    |v_{k'}^{({\rm R})}\rangle\delta_{jj'} 
  -\langle v_k^{({\rm R})}|\phi_{n+1}
    |v_{k'}^{({\rm R})}\rangle
    \langle j^{(n)}|\phi_n|j'^{(n)}\rangle
  +\frac{1}{2}\langle j^{(n)}|\phi_n^2|j'^{(n)}\rangle\delta_{kk'}. 
\end{eqnarray*}
To calculate the matrix elements (\ref{h}), 
the following matrix elements need to be calculated and stored 
at every DMRG step. 
\begin{eqnarray*}
&\langle u_i^{({\rm L})}|\bar{H}_{\rm L}|u_{i'}^{({\rm L})}\rangle,\quad
\langle v_k^{({\rm R})}|\bar{H}_{\rm R}|v_{k'}^{({\rm R})}\rangle,&\\
&\langle u_i^{({\rm L})}|\phi_{n-1}^2|u_{i'}^{({\rm L})}\rangle,\quad
\langle v_k^{({\rm R})}|\phi_{n+1}^2|v_{k'}^{({\rm R})}\rangle,&\\
&\langle u_i^{({\rm L})}|\phi_{n-1}|u_{i'}^{({\rm L})}\rangle,\quad
\langle v_k^{({\rm R})}|\phi_{n+1}|v_{k'}^{({\rm R})}\rangle.&
\end{eqnarray*}
Other matrix elements do not need to be stored 
because they can be calculated with operator contraction. 
(See Appendix \ref{me}.)

\section{Numerical analysis}
\label{result}
The critical values are defined 
with vacuum expectation values of the variable $\phi_n$: 
\begin{equation}
  v(\tilde{\lambda},\tilde{\mu}^2)\equiv
  \langle \Psi_0|\phi_n|\Psi_0\rangle=
  A(\tilde{\lambda})\left[
    \frac{\tilde{\lambda}}{\tilde{\mu}^2}-
    \frac{\tilde{\lambda}}{\tilde{\mu}_{\rm c}^2
    (\tilde{\lambda})}
  \right]^{\beta(\tilde{\lambda})}, 
  \label{v}
\end{equation}
where $|\Psi_0\rangle$ is a vacuum state and 
$A(\tilde{\lambda})$ is a constant dependent on $\tilde{\lambda}$. 
The vacuum expectation value $v$ is calculated at the center $n=L/2$ 
using vacuum wavefunctions $\Psi_{ijk}$
for each pair of $\tilde{\lambda}$ and $\tilde{\mu}^2$. 
The quantities $\tilde{\mu}_{\rm c}^2(\tilde{\lambda})$, 
$\beta(\tilde{\lambda})$, and $A(\tilde{\lambda})$ 
are determined by fitting $v$ to obtained numerical data. 
The critical coupling $(\lambda/\mu^2)_{\rm c}$ 
and the critical exponent $\beta$ are defined 
in the continuum limit $a\to 0$ as follows \cite{Loinaz:1997az}: 
\begin{equation}
  \displaystyle\frac{\tilde{\lambda}}{\tilde{\mu}_{\rm c}^2
  (\tilde{\lambda})}=
  \left(\frac{\lambda}{\mu^2}\right)_{\rm c}
  +B\tilde{\lambda},
  \label{ext1}
\end{equation}
\begin{equation}
  \beta(\tilde{\lambda})=\beta+C\tilde{\lambda}, 
  \label{ext2}
\end{equation}
where $B$ and $C$ are some constants. 
In the continuum limit $a\to 0$, 
the coupling constant $\tilde{\lambda}=\lambda a^2$ vanishes. 
Then, we have 
\begin{equation}
  \left(\frac{\lambda}{\mu^2}\right)_{\rm c}=
  \lim_{\tilde{\lambda}\to 0}
  \frac{\tilde{\lambda}}{\tilde{\mu}_{\rm c}^2(\lambda)}, \quad
  \beta=\lim_{\tilde{\lambda}\to 0}\beta(\tilde{\lambda}). 
\end{equation}
In numerical analysis, the continuum limit cannot be 
reached directly. 
The critical values $(\lambda/\mu^2)_{\rm c}$ and 
$\beta$ are determined by extrapolating 
$\tilde{\lambda}/\tilde{\mu}_{\rm c}^2(\tilde{\lambda})$ 
and $\beta(\tilde{\lambda})$ 
to the continuum limit $\tilde{\lambda}=0$. 
Also, finite-size effects on the critical values are removed 
by using sufficiently large lattices ($L=250,500$, and $1000$). 

Figure \ref{conv_e_mn} shows $N$ dependence of vacuum energy 
$\tilde{E}_0=E_0 a$ for each $M=8,9$, and $10$. 
Values of the other parameters are $\tilde{\lambda}=0.6$, 
$\tilde{\mu}_0^2=-0.2$ ($\tilde{\mu}^2=9.744828\times 10^{-3}$), 
and $L=1000$. 
The energy converges as $N$ increases with $M$ fixed. 
Parameter values $(M,N)=(10,10)$ gives convergence of vacuum energy 
in accuracy of four digits. 
Figure \ref{conv_v_mn} shows $N$ dependence of the vacuum 
expectation value $v$ for each $M=8,9$, and $10$. 
Other parameter values are same as Fig.~\ref{conv_e_mn}. 
$v$ converges at $N=10$ for each $M$, 
but convergence for $M$ with $N$ fixed is slow. 
We will use the values $(M,N)=(10,10)$ 
in all calculations to determine critical values. 

\FIGURE{
  \epsfig{file=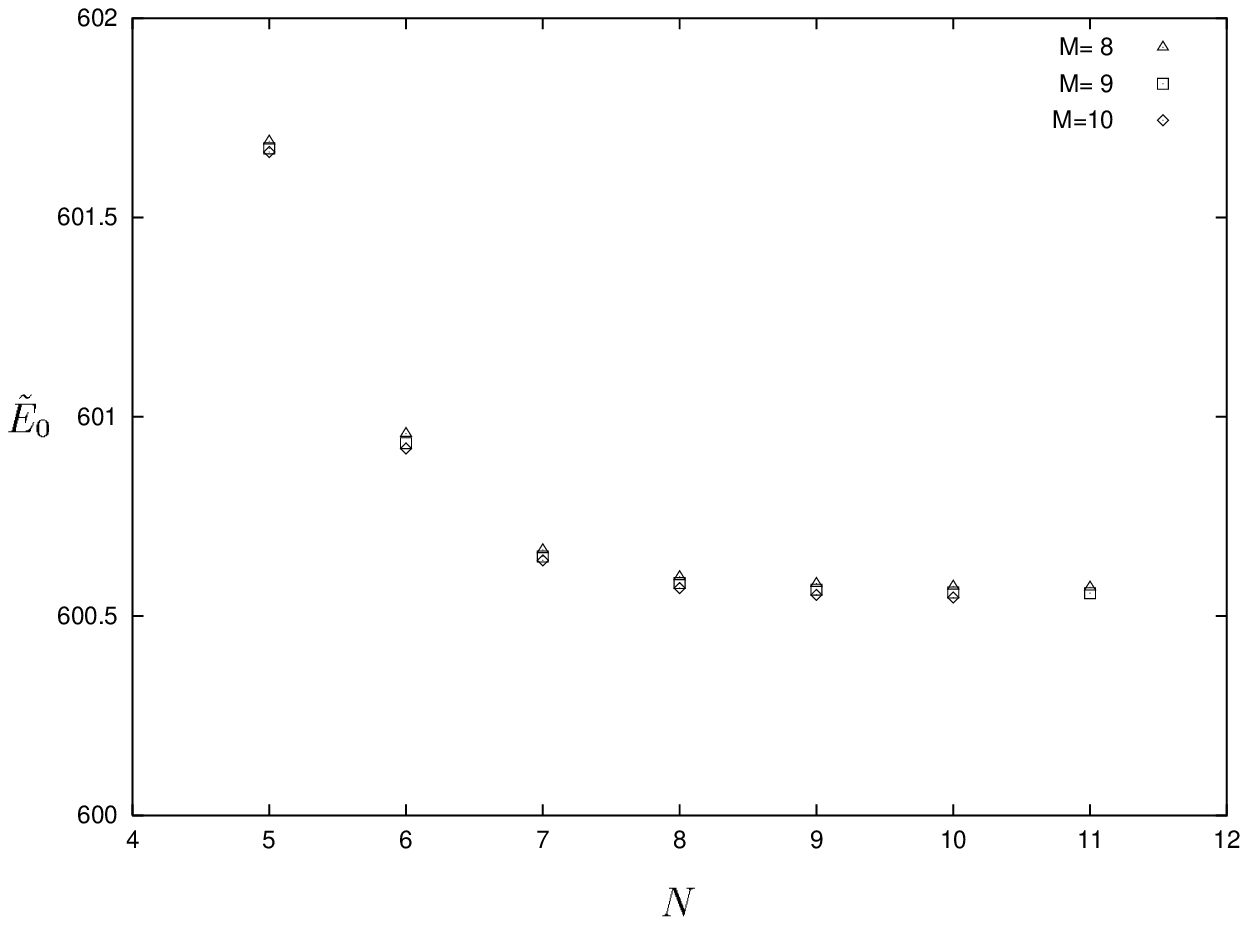,width=10.0cm}
\caption{
Vacuum energy $\tilde{E}_0=E_0 a$ is plotted as a function of $N$ 
for $M=8,9$, and $10$. 
Coupling constant, bare mass squared, and lattice size are 
$\tilde{\lambda}=0.6$, $\tilde{\mu}_0^2=-0.2$, 
and $L=1000$, respectively.}
\label{conv_e_mn}}

\FIGURE{
  \epsfig{file=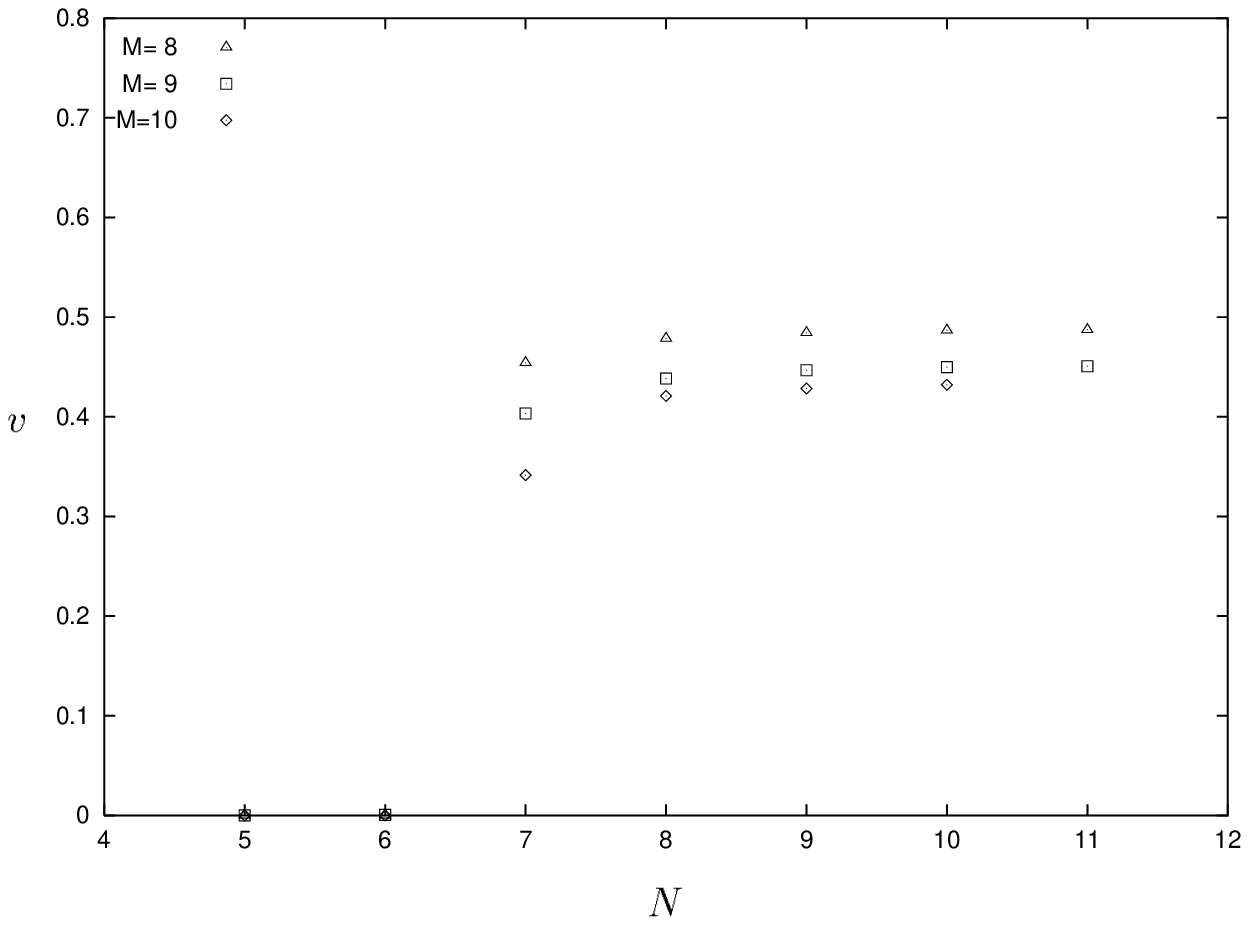,width=10.0cm}
\caption{
Vacuum expectation value $v$ is plotted as a function of $N$ 
for $M=8,9$, and $10$. 
Coupling constant, bare mass squared, and lattice size are 
$\tilde{\lambda}=0.6$, $\tilde{\mu}_0^2=-0.2$, 
and $L=1000$, respectively. }
\label{conv_v_mn}}

\FIGURE{
 \epsfig{file=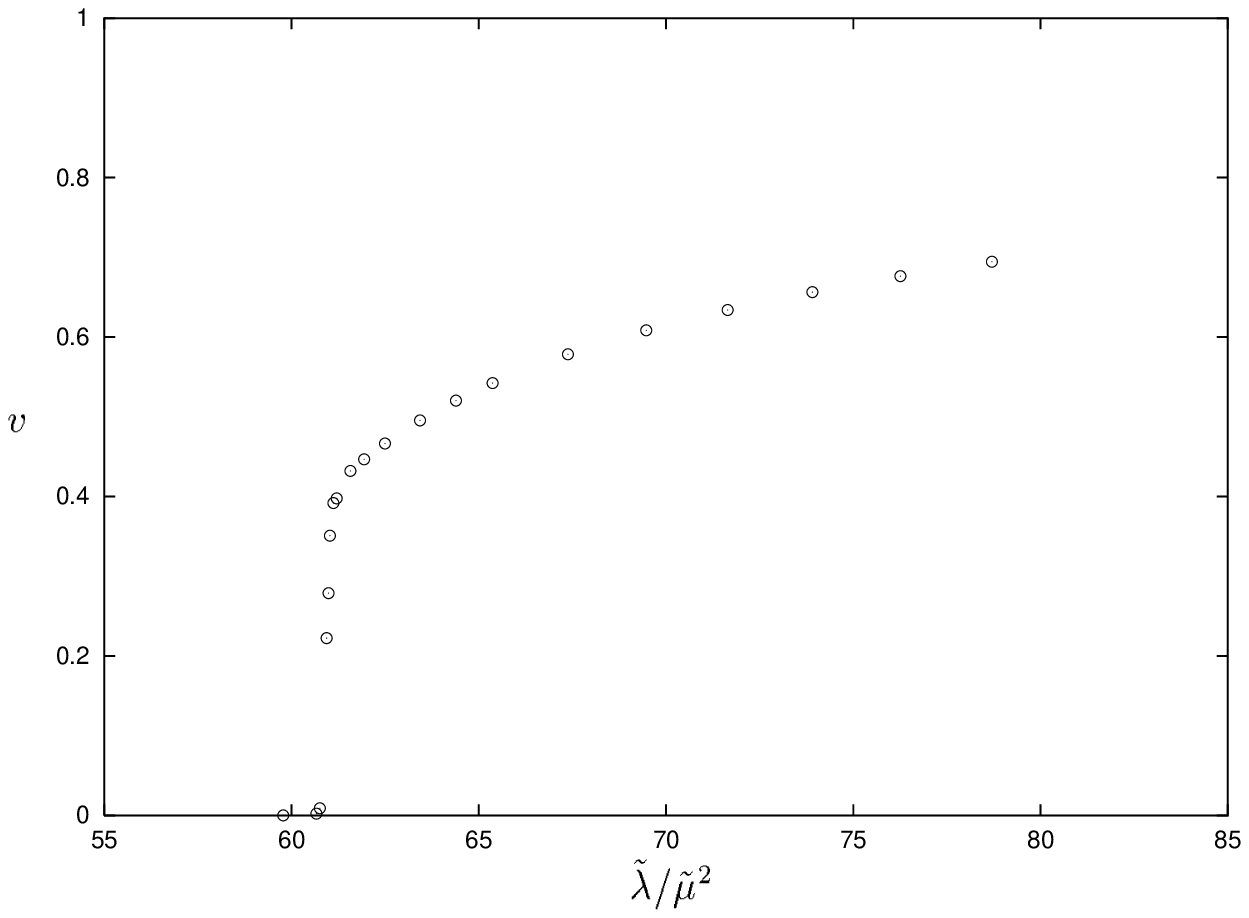,width=10.0cm}
\caption{Vacuum expectation value $v$ is plotted 
as a function of $\tilde{\lambda}/\tilde{\mu}^2$ 
with $\tilde{\lambda}=0.6$ fixed. 
The lattice size is $L=1000$. 
}
\label{vev}}

In Fig. \ref{vev}, vacuum expectation value 
$v(\tilde{\lambda},\tilde{\mu}^2)$ 
is plotted as a function of $\tilde{\lambda}/\tilde{\mu}^2$ 
for a fixed $\tilde{\lambda}=0.6$ on the lattice $L=1000$. 
If $\tilde{\lambda}/\tilde{\mu}^2$ is decreased from the positive 
side, $v$ becomes smaller and finally vanishes at a point 
$\tilde{\lambda}/\tilde{\mu}_{\rm c}^2(\tilde{\lambda})$. 
The obtained data points are fitted with Eq. (\ref{v}) 
to calculate $\tilde{\mu}_{\rm c}^2(\tilde{\lambda})$ and 
$\beta(\tilde{\lambda})$ for each $\tilde{\lambda}$. 
The Marquardt-Levenberg nonlinear least-squares algorithm 
is used for fitting. 
Similar figures are drawn also for $\tilde{\lambda}=1.5$ and $3.0$ 
to determine $\tilde{\mu}_{\rm c}^2(\tilde{\lambda})$ and 
$\beta(\tilde{\lambda})$. 

\FIGURE{
\epsfig{file=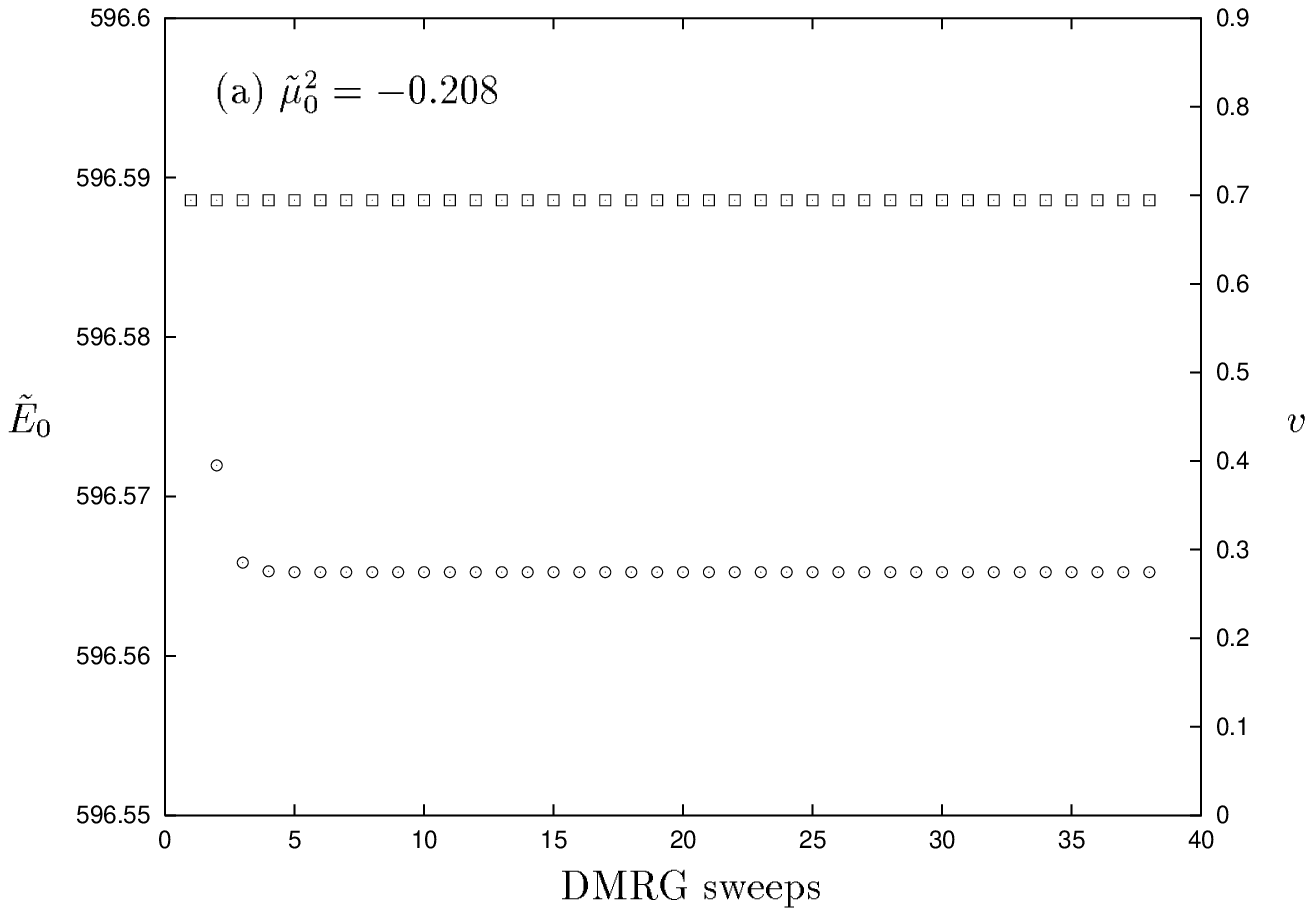,width=7.0cm}
\epsfig{file=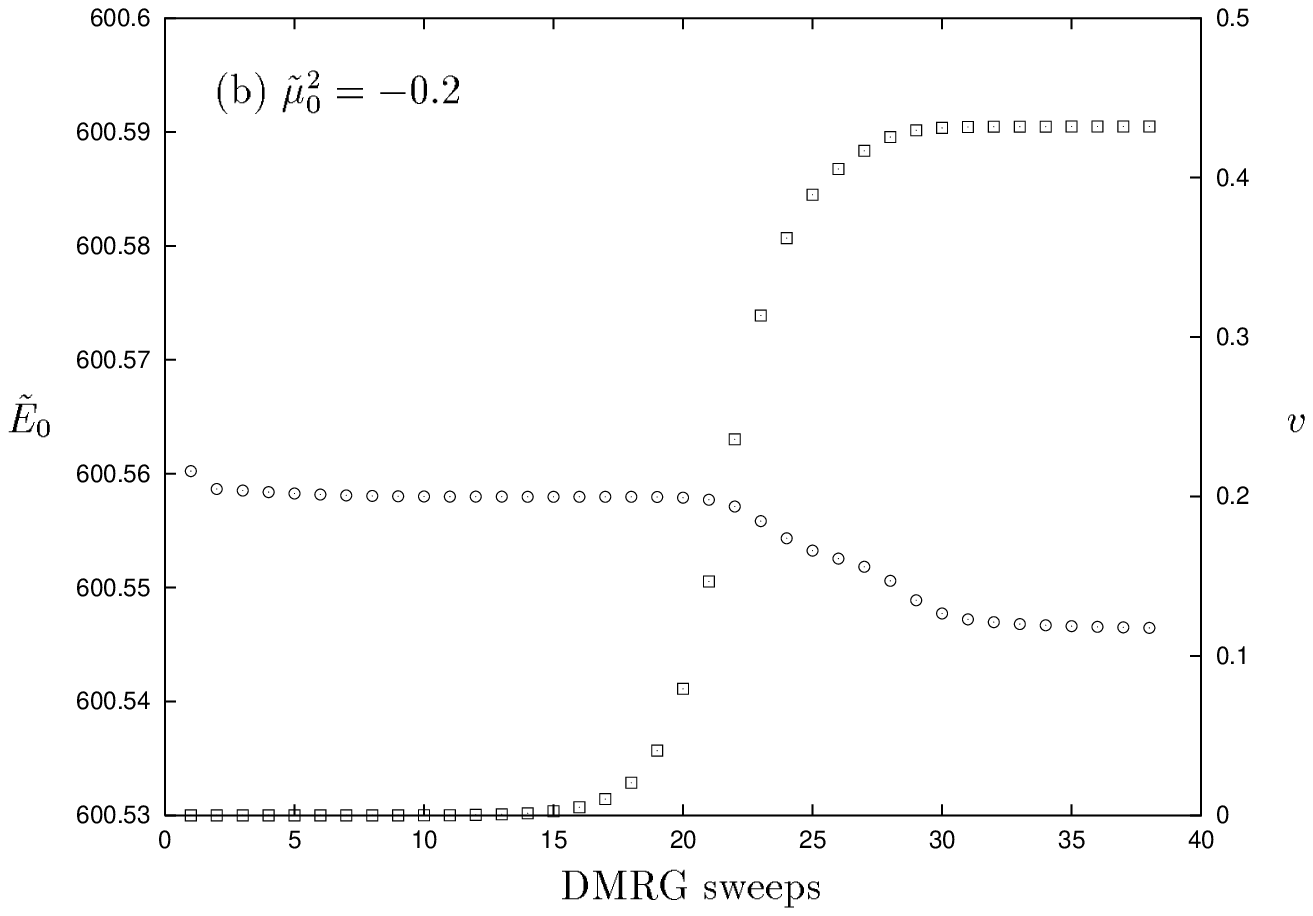,width=7.0cm}
\caption{$\tilde{E}_0$ (circles) and $v$ (squares)
are plotted as functions of the number of 
DMRG sweeps for $\tilde{\lambda}=0.6$, $L=1000$, and $(M,N)=(10,10)$. 
Convergence speed are compared between 
(a) $\tilde{\mu}_0^2=-0.208$ 
($\tilde{\lambda}\tilde{\mu}^2\sim 78.7$) and 
(b) $\tilde{\mu}_0^2=-0.2$ 
($\tilde{\lambda}\tilde{\mu}^2\sim 61.6$). 
Circles and squares represent $\tilde{E}_0$ and $v$, respectively. 
Data points are plotted every DMRG sweep. 
}
\label{conv_rg_1}}

In Fig. \ref{conv_rg_1},  $\tilde{E}_0$ and $v$ are plotted 
as functions of the number of DMRG sweeps 
for $\tilde{\lambda}=0.6$ and $L=1000$. 
Convergence speed is compared between $\tilde{\mu}_0^2=-0.208$ 
($\tilde{\lambda}/\tilde{\mu}^2\sim 78.7$) and
$\tilde{\mu}_0^2=-0.2$ ($\tilde{\lambda}/\tilde{\mu}^2\sim 61.6$). 
The former and latter parameter values correspond to 
the first and twelfth points from the right hand side 
in Fig. \ref{vev}, respectively. 
The latter is close to the critical value 
$\tilde{\lambda}/\tilde{\mu}_{\rm c}^2(\tilde{\lambda})|
_{\tilde{\lambda}=0.6}$. 
In Fig. \ref{conv_rg_1},  
data points are plotted every DMRG sweep. One DMRG sweep is 
composed of $L=1000$ sets of basis state optimization. 
Convergence tends to be slow as $\tilde{\lambda}/\tilde{\mu}^2$ 
approaches the critical value. 
When $\tilde{\mu}_0^2=-0.208$, convergence of 
both quantities is very fast. On the other hand, 
when $\tilde{\mu}_0^2=-0.2$, energy converges in accuracy of 
four digits even with a small number of 
DMRG sweeps, but convergence of $v$ is slow. 
$v$ barely converges in the latter case. 
Because of limited computer time, 
complete convergence is not available for $v$ when 
$\tilde{\mu}^2$ is very close to the critical value 
$\tilde{\mu}_{\rm c}^2(\tilde{\lambda})$. 
We calculate $\tilde{E}_0$ and $v$ with up to forty DMRG sweeps 
checking their convergence to determine 
$\tilde{\mu}_{\rm c}^2(\tilde{\lambda})$ on each lattice. 

\FIGURE{
  \epsfig{file=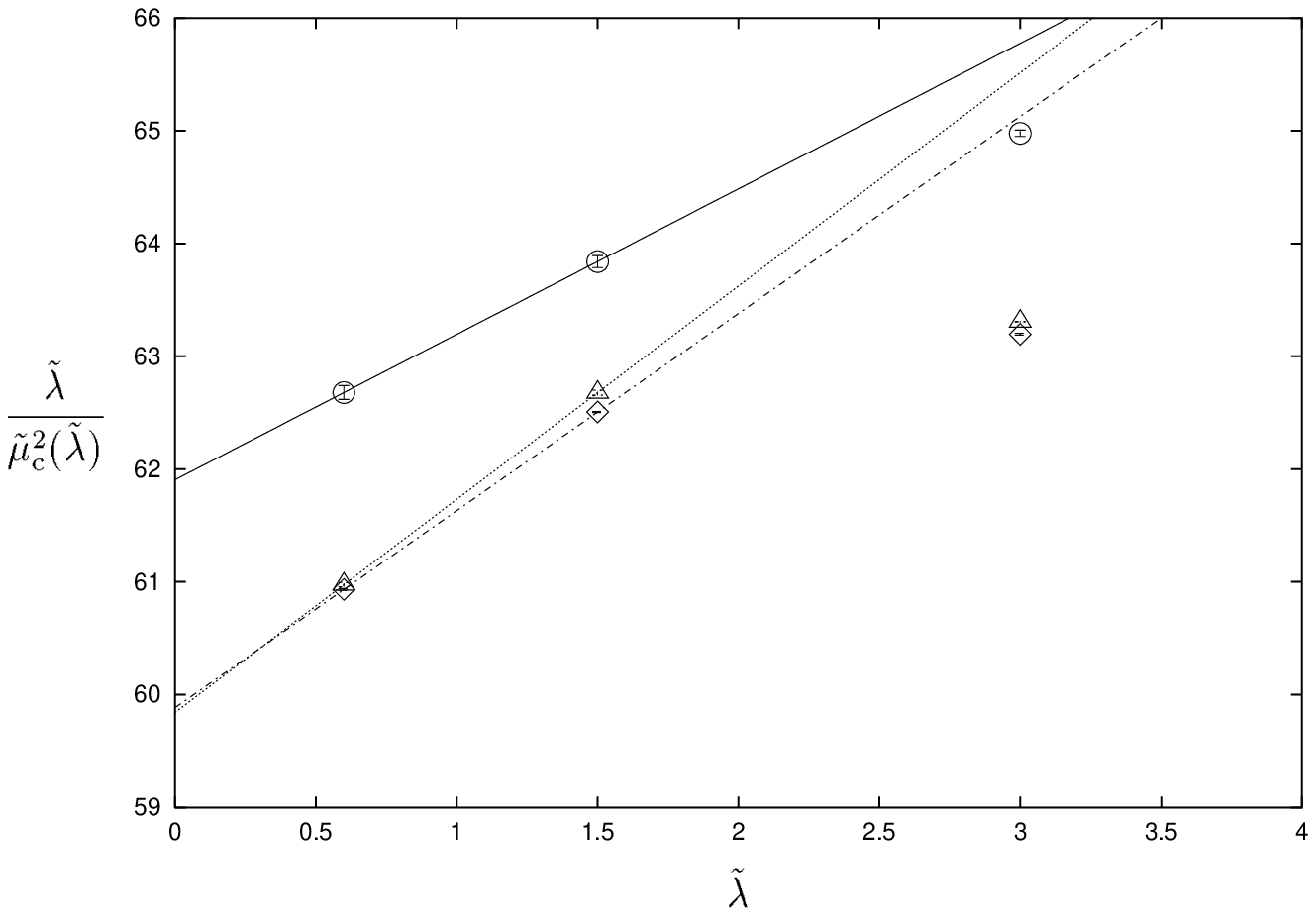,width=10.0cm}
\caption{The ratio 
$\tilde{\lambda}/\tilde{\mu}^2_{\rm c}(\tilde{\lambda})$ 
is plotted as a function of the coupling constant 
$\tilde{\lambda}$ ($=0.6,1.5,3.0$) for three lattice sizes 
$L=250$ (circles), $500$ (triangles), and $1000$ (diamonds). 
Each plot is fitted with a straight line to determine 
$(\lambda/\mu^2)_{\rm c}$ in the limit $\tilde{\lambda}\to 0$. 
The fitted lines are drawn with solid ($L=250$), 
dotted ($L=500$), and dot-dashed ($L=1000$) lines. 
}
\label{cc}}

\FIGURE{
  \epsfig{file=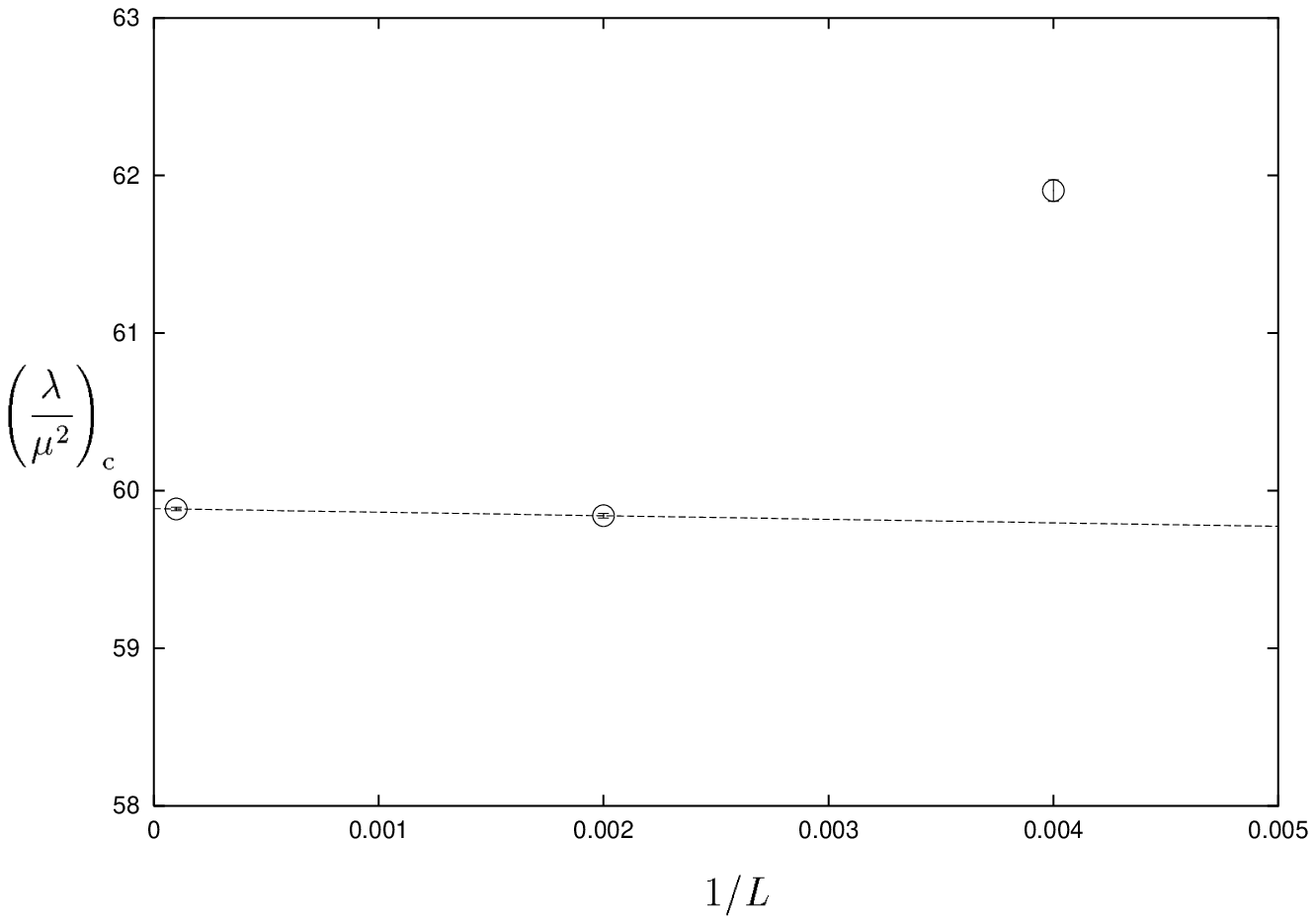,width=10.0cm}
\caption{The critical value $(\lambda/\mu^2)_{\rm c}$
is plotted as a function of $1/L$
for $L=250,500$, and $1000$. 
Extrapolation to the limit $L\to\infty$ gives 
$(\lambda/\mu^2)_{\rm c}=59.89\pm 0.01$. }
\label{cc_cntlim}}

In Fig. \ref{cc}, the ratio 
$\tilde{\lambda}/\tilde{\mu}^2_{\rm c}(\tilde{\lambda})$ is 
plotted as a function of $\tilde{\lambda}$ ($=0.6,1.5,3.0$) 
with error bars for three lattice sizes $L=250$, $500$, and $1000$. 
The errors come from fitting. 
We are going to take the continuum limit $a\to 0$ 
to determine the values of $(\lambda/\mu^2)_{\rm c}$. 
As explained in Eq. (\ref{ext1}), 
the data points are fitted with a straight line 
and extrapolated to the limit $\tilde{\lambda}\to 0$ for each $L$. 
The data points for $\tilde{\lambda}=0.6$ and $1.5$ are 
used for fitting and extrapolation. 

Figure \ref{cc_cntlim} plots the extrapolated values 
$(\lambda/\mu^2)_{\rm c}$ as a function of $1/L$. 
We observe good agreement between the two results for 
$L=500$ and $1000$. 
That is, the lattice size $L=500$ is sufficiently large and 
close to the limit $L\to\infty$. 
For this reason, we fit the two points for $L=500$ and $1000$ 
with a straight line to extrapolate them to the limit $L\to\infty$. 
In the continuum limit $a\to 0$ and $L\to\infty$, 
we obtain $(\lambda/\mu^2)_{\rm c}=59.89\pm 0.01$. 
Our result is close to the Euclidean Monte Carlo result 
\cite{Loinaz:1997az}, 
which has been given with lattices up to $L=512$. 
No exact result for $(\lambda/\mu^2)_{\rm c}$ has been known. 
Table \ref{ccc} shows various results 
for the critical value $(\lambda/\mu^2)_{\rm c}$. 

\TABLE{
\caption{Various results for the critical coupling constant 
$(\lambda/\mu^2)_{\rm c}$ are listed. 
Our result is consistent with Monte Carlo within the errors. } 
\label{ccc}
\begin{tabular}{lcc}
\hline
Method & Result & Reference\\
\hline
DMRG                         & $59.89\pm 0.01$ & This work\\
Monte Carlo                  & $61.56_{-0.24}^{+0.48}$ & \cite{Loinaz:1997az}\\
Gaussian effective potential & $61.266$ & \cite{chang}\\
Gaussian effective potential & $61.632$ & \cite{Hauser:mb}\\
Connected Green function     & $58.704$ & \cite{Hauser:mb}\\
Coupled cluster expansion    & $22.8<(\lambda/\mu^2)_{\rm c}<51.6$ 
& \cite{Funke:wb}\\
Non-Gaussian variational     & $41.28$ & \cite{Polley:wf}\\
Discretized light cone       & $43.896$, $33.000$ &
 \cite{Harindranath:db,Harindranath:zt}\\
Discretized light cone       & $42.948$, $46.26$ &
 \cite{Sugihara:1997xh}\\ \hline
\end{tabular}
}

\FIGURE{
 \epsfig{file=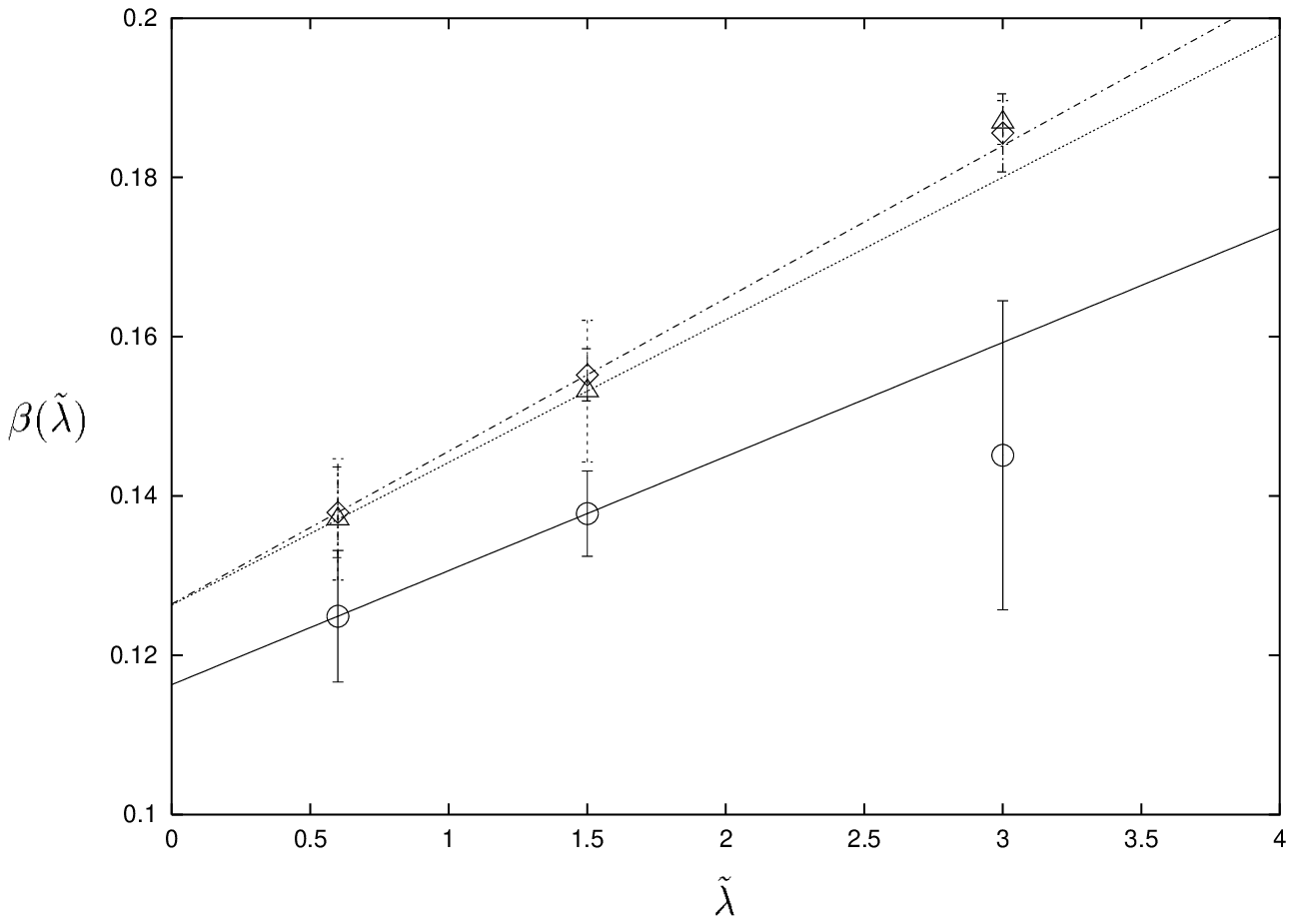,width=10.0cm}
\caption{$\beta(\tilde{\lambda})$ 
is plotted as a function of the coupling constant $\tilde{\lambda}$ 
($=0.6,1.5,3.0$) for three lattice sizes 
$L=250$ (circles), $500$ (triangles), and $1000$ (diamonds). 
Each plot is fitted with a straight line to determine $\beta(0)$. 
Fitted lines are drawn with solid (L=250), 
dotted ($L=500$), and dot-dashed ($L=1000$) lines.
}
\label{ce}}

\FIGURE{
 \epsfig{file=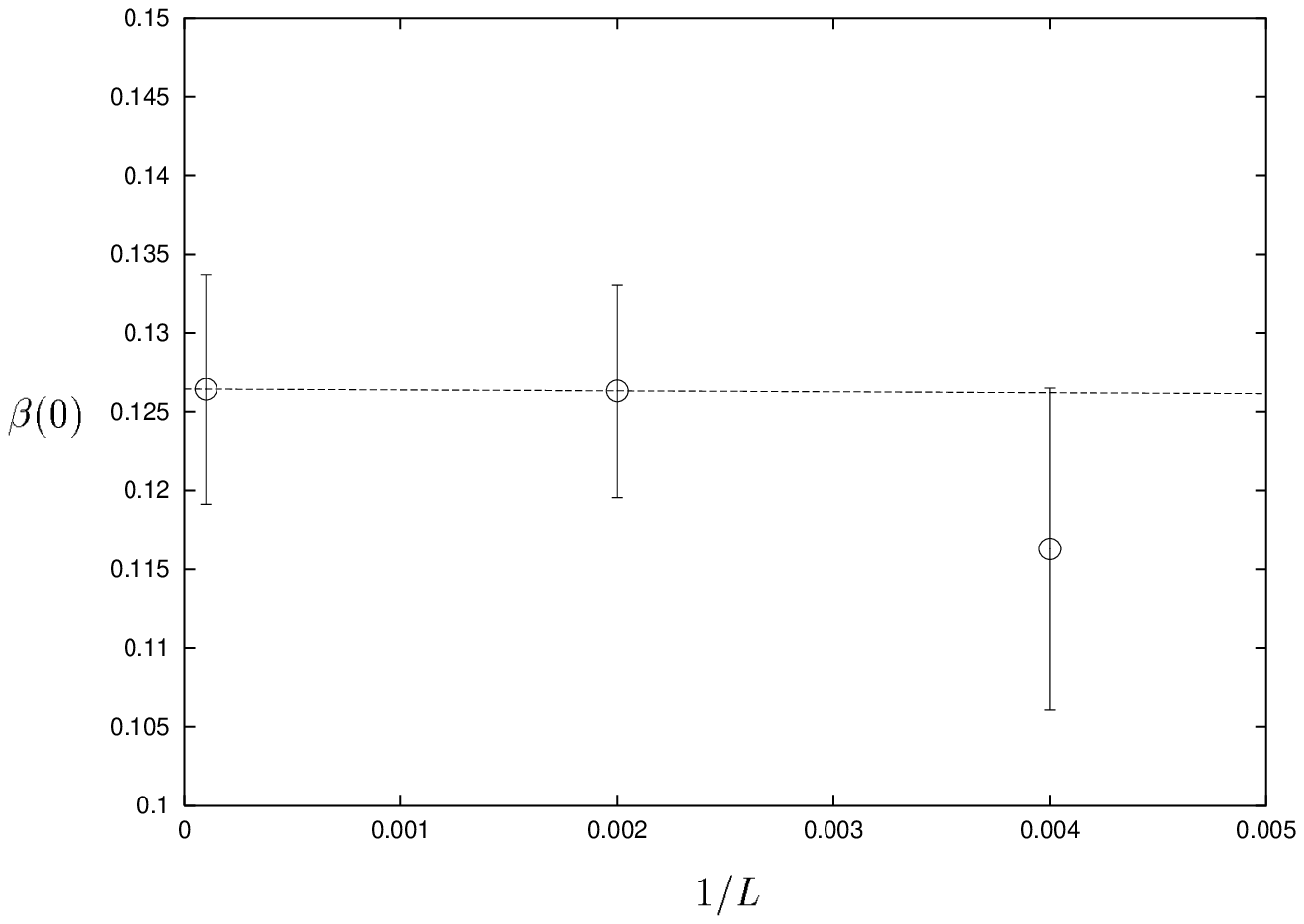,width=10.0cm}
\caption{$\beta(0)$ is plotted as a function of $1/L$ 
for $L=250,500$, and $1000$. 
Extrapolation to the limit $L\to\infty$ 
gives $\beta=0.1264\pm 0.0073$, 
which is consistent with the exact result $\beta=1/8=0.125$. }
\label{ce_cntlim}}

Figure \ref{ce} plots the critical exponent $\beta(\tilde{\lambda})$ 
as a function of $\tilde{\lambda}$ ($=0.6,1.5,3.0$) 
for $L=250$, $500$, and $1000$. 
As before, the data points are fitted with Eq. (\ref{ext2}) and 
extrapolated to the limit $\tilde{\lambda}=0$ for each $L$. 
The data points for $\tilde{\lambda}=0.6$ and $1.5$ are 
used for fitting and extrapolation. 
Figure \ref{ce_cntlim} plots the extrapolated values $\beta(0)$ 
as a function of $1/L$. 
Also in this case, we observe that the results 
for $L=500$ and $1000$ are very close. 
Extrapolation with these two points to the limit $L\to\infty$ 
gives the critical exponent $\beta=0.1264\pm 0.0073$. 
No Monte Carlo result for the critical exponent $\beta$ 
is available in the literature at this point. 
According to the universality class consideration, 
the critical exponents of the (1+1)-dimensional 
$\lambda\phi^4$ model are same as the two-dimensional Ising model, 
which has been exactly solved by Onsager \cite{onsager}. 
Our result is consistent with the exact value $\beta=1/8=0.125$. 
As seen from Fig. \ref{cc_cntlim} and \ref{ce_cntlim}, the lattice size 
$L=500$ is sufficiently close to the limit $L\to\infty$.

\section{Summary}
\label{summary}
We have studied spontaneous breakdown of $Z_2$ symmetry 
of a (1+1)-dimensional $\phi^4$ model 
using the density matrix renormalization group technique. 
We have determined the critical coupling constant 
$(\lambda/\mu^2)_{\rm c}$ and the critical exponent $\beta$ 
of the model by extrapolating the numerical results for finite 
lattices to the continuum limit $a\to 0$ and $L\to\infty$. 
DMRG truncation works well also in the bosonic model. 
The lattice with $L=500$ can give results sufficiently 
close to the limit $L\to\infty$. 
To improve calculations near $\tilde{\mu}^2_{\rm c}(\tilde{\lambda})$, 
we need to find a way to include effects of 
large quantum fluctuations.

\acknowledgments

The numerical calculations were carried on 
RIKEN VPP and  Yukawa Institute SX5. 
This work has been partially supported by RIKEN. 

\appendix
\section{Basis states and density matrices}
\label{appa}
This is a brief explanation of basis state creation 
with the density matrix technique, 
which is the key ingredient of DMRG. 
We split a spatial lattice into left and right blocks, 
each of which are expressed with basis sets 
$\{|i^{({\rm L})}\rangle|i=1,2,\dots,N_{\rm L}\}$ and 
$\{|j^{({\rm R})}\rangle|j=1,2,\dots,N_{\rm R}\}$, respectively. 
The target state to be solved is expanded as 
\begin{equation}
  |\Psi\rangle = \sum_{i=1}^{N_{\rm L}} \sum_{j=1}^{N_{\rm R}}
  \Psi_{ij}|i^{({\rm L})}\rangle |j^{({\rm R})}\rangle, 
  \quad
  N_{\rm L}\le N_{\rm R}. 
\end{equation}
When $N_{\rm L}$ and $N_{\rm R}$ are sufficiently small, 
we can obtain wavefunction of a target state 
(say ground state) by diagonalizing Hamiltonian numerically. 
If the basis states 
$\{|i^{({\rm L})}\rangle\}$ and $\{|j^{({\rm R})}\rangle\}$ 
are transformed, 
the wavefunction $\Psi_{ij}$ is also transformed. 
We want to make the absolute values of the wavefunction components 
very small as many as possible, because basis states giving very small 
wavefunction component are not important for 
the target state $|\Psi\rangle$ and can be thrown away. 
We transform basis states based on 
the following singular-value decomposition. 
\begin{equation}
  \Psi_{ij} = \sum_{k=1}^{N_{\rm L}} U_{ik}D_k V_{kj}, 
\end{equation}
where
\[
  \sum_{i=1}^{N_{\rm L}} U_{ik}^* U_{ik'}=\delta_{kk'}, \quad
  \sum_{j=1}^{N_{\rm R}} V_{kj}^* V_{k'j}=\delta_{kk'}. 
\]
In the new basis 
\[
  |u_k^{({\rm L})}\rangle=\sum_{i=1}^{N_{\rm L}} U_{ik}|i^{({\rm L})}\rangle, 
  \quad
  |v_k^{({\rm R})}\rangle=\sum_{j=1}^{N_{\rm R}} V_{kj}|j^{({\rm R})}\rangle, 
\]
the target state becomes
\begin{equation}
  |\Psi\rangle = \sum_{k=1}^{N_{\rm L}} D_k
  |u_k^{({\rm L})}\rangle |v_k^{({\rm R})}\rangle. 
\end{equation}
The result does not change largely even if 
a basis state with small $D_k$ is removed from the calculation. 
On the other hand, a basis state with large $D_k$ is important 
and cannot be neglected. 
We can use $D_k$'s to choose good basis states and 
control calculation accuracy. 
In actual numerical works, we diagonalize the following 
density matrices to obtain $U$, $V$, and $D$ 
in stead of performing singular-value decomposition directly. 
\begin{eqnarray*}
  \rho_{ii'}^{({\rm L})}&=&\sum_{j=1}^{N_{\rm R}} \Psi_{ij}^* \Psi_{i'j}
  = \sum_{k=1}^{N_{\rm L}} U_{ik}^* |D_k|^2 U_{i'k}, 
  \\
  \rho_{jj'}^{({\rm R})}&=&\sum_{i=1}^{N_{\rm L}} \Psi_{ij}^* \Psi_{ij'}
  = \sum_{k=1}^{N_{\rm L}} V_{kj}^* |D_k|^2 V_{kj'}. 
\end{eqnarray*}
See Ref. \cite{dmrglec} for the details of DMRG 
and related topics. 

\section{Matrix elements}
\label{me}
This appendix gives formulas for matrix elements 
of powers of the field operators $\pi_n$ and $\phi_n$. 
For the definition of the basis states $|j^{(n)}\rangle$, 
see Eq. (\ref{state}). 
\begin{eqnarray*}
  \langle j^{(n)}|\pi_n^2|j'^{(n)}\rangle
  &=&\frac{1}{2}
  \bigg[-\sqrt{(j-1)(j-2)}\delta_{j-1,j'+1}
  \nonumber
  \\
  && +(2j-1)\delta_{j,j'}
  -\sqrt{(j'-1)(j'-2)}\delta_{j+1,j'-1}\bigg], 
  \\
  \langle j^{(n)}|\phi_n|j'^{(n)}\rangle
  &=&
  \frac{1}{\sqrt{2}}
  \bigg[
  \sqrt{j-1}\delta_{j-1,j'}+\sqrt{j'-1}\delta_{j,j'-1}
  \bigg],
  \\
  \langle j^{(n)}|\phi_n^2|j'^{(n)}\rangle 
  &=&
  \frac{1}{2}
  \bigg[
  \sqrt{(j-1)(j-2)}\delta_{j-1,j'+1}
  \nonumber
  \\
  && +(2j-1)\delta_{j,j'}
  +\sqrt{(j'-1)(j'-2)}\delta_{j+1,j'-1}\bigg], 
  \\
  \langle j^{(n)}|\phi_n^4|j'^{(n)}\rangle&=&
  \frac{1}{4}
  \bigg[
  \sqrt{(j-1)(j-2)(j-3)(j-4)}\delta_{j-1,j'+3}
  \nonumber
  \\
  && +4j\sqrt{(j-1)(j-2)}\delta_{j,j'+2}
  \nonumber
  \\
  && -6\sqrt{(j-1)(j-2)}\delta_{j-1,j'+1}
  \nonumber
  \\
  && +3(2j^2-2j+1)\delta_{j,j'}
  \nonumber
  \\
  && + 4j'\sqrt{(j'-1)(j'-2)}\delta_{j+2,j'}
  \nonumber
  \\
  && -6\sqrt{(j'-1)(j'-2)}\delta_{j+1,j'-1}
  \nonumber
  \\
  && +\sqrt{(j'-1)(j'-2)(j'-3)(j'-4)}\delta_{j+3,j'-1}
  \bigg]. 
\end{eqnarray*}

\end{document}